# Magnetization, thermoelectric, and pressure studies of the magnetic field-induced metal to insulator transition in tau phase organic conductors

D. Graf, E. S. Choi, and J. S. Brooks

NHMFL/Physics, Florida State University, Tallahassee, FL 32310, USA

N. Harrison

NHMFL/Los Alamos National Laboratory, Los Alamos, NM 87545, USA

K. Murata

Graduate School of Science, Osaka City University, Sumiyoshi-ku, Osaka 558-8585, Japan

T. Konoike

National Institute for Materials Science, Nanomaterials Laboratory, 3-13, Sakura, Tsukuba, Ibaraki 305-0003, Japan

G. Papavassiliou

Theoretical and Physical Chemistry Institute, National Hellenic Research Foundation, Athens, 116-35, Greece

**Abstract**

We have investigated the magnetic field-induced metal-insulator transition in the τ-phase organic conductors, which occurs in fields above 35 T, and below 14 K, by magnetization, thermoelectric, and pressure dependent transport methods. Our results show that the transition is a bulk thermodynamic process where a magnetic field-dependent gap opens upon entry into the insulating state. We argue that the transition involves a magnetic field-induced change in the electronic structure.

**Introduction**

We report magnetization, thermoelectric and pressure dependent transport measurements of the magnetic field-induced metal to insulator (MI) transition in the τ-phase organic conductors τ-(P-(S,S)-DMEDT-TTF)$_2$(AuBr$_2$)$_{1+y}$, τ-(P-(r)-DMEDT-TTF)$_2$(AuBr$_2$)$_{1+y}$, and τ-(P-($S,S$)-DMEDT-TTF)$_2$(AuI$_2$)$_{1+y}$. The MI transition has been previously reported as a rapid rise in the resistance from a metal to an insulator in the range 30 to 50 T at temperatures below 14 K. [1,2] The MI transition is hysteretic in field, and is also observed in magnetocaloric and skin-depth measurements. The threshold field for the MI transition ($B_{MI}$) is found to be nearly independent of field direction and may be driven by isotropic spin rather than Fermi surface (FS) nesting effects. Although a metallic state with Shubnikov-de Haas (SdH) oscillations is often observed below the threshold field, the resistance above the threshold field in the insulating state is immeasurable.

For τ-(P-(S,S)-DMEDT-TTF)$_2$(AuBr$_2$)$_{1+y}$, P (i.e. pyrazino) represents the DMEDT-TTF organic cation with a N-N end group. The AuBr$_2$ linear anions reside stoichiometrically (2:1) in the square network of cations in the *ab*-conducting planes, and also non-stoichiometrically (y ~ 0.75) in between these molecular layers. Due to the low symmetry of the donor, and the additional interplanar anion arrangement, the unit cell is large and involves four donor layers, with crystallographic axes *a*, *b,* and *c* ≈ 7.3, 7.3 and 68 Å, respectively. Materials containing one type of isomer of the donor molecule are noted by (S,S) or (R,R) in the stoichiometry while materials with more than one type of isomer are racemic, noted by (*r*) [see refs. 3-5 for a more detailed view of the crystal structure]. We refer to the pyrazino (S,S) and racemic samples as τ-AuBr$_2$ , τ-AuI$_2$ , and τ(*r*)-AuBr$_2$, respectively. The same structure where the N-N group is replaced with an O-O oxygen group is τ-(EDO-($S,S$)-DMEDT-TTF)$_2$(AuBr$_2$)$_{1+y}$ (hereafter, τ-EDO-AuBr$_2$). The EDO material does not show a pronounced MI transition, and differs from the pyrazino compounds in other aspects, as discussed below. The ground state below 36 T in this class of materials has been studied by magnetoresistance (MR), magnetization, and pressure. [6,7] Many anomalies arise that suggest magnetic moments and magnetic memory effects occur.

The tight binding band structure and FS is shown in Fig. 1, based on refined crystallographic studies at 290 K. An unusual feature of the electron band is the narrow, nearly dispersionless appearance. (Arita *et al*. have considered the possibility that the very flat band may lead to the apparent magnetic properties, including the pronounced negative MR at low fields. [8]) The resulting FS has a four-fold symmetry with a single SdH frequency of about 900 T. The in-plane angular dependent MR exhibits a four-fold symmetry in high fields[9]. However, there are multiple SdH oscillations observed[10,11], and the highest frequency ($F_3$) is only ~ 500 T. Moreover, although $F_3$ is seen in both the pyrazino and EDO compounds, a lower frequency, $F_2 = 180$ T, with smaller amplitude than $F_3$ is observed for pyrazino. Three frequencies, including the lowest, $F_1 = 50$ T, are seen for τ-EDO-AuBr$_2$ where $F_1$ has an amplitude much larger than $F_2$ or $F_3$ in the FFT spectrum.

As discussed above, the MI transition appears to be driven by an isotropic spin (Zeeman) effect, but the temperature dependence of the phase diagram is contrary to the behavior of a standard charge density wave ground state which is removed at the Pauli limit. The transition resembles a field-induced spin-density wave state ($T_{MI}$ increases with field), but the transition does not follow conventional $1/\cos(\theta)$ behavior. The magnetocaloric data was highly suggestive of a thermodynamic mechanism[1], but the non-equilibrium nature of the measurement left open questions about eddy current effects and further experiments were necessary to characterize the bulk, thermodynamic nature of the MI transition. For this, we have undertaken magnetization, magnetothermopower, and pressure studies.

**Experiment**

Samples were grown electrochemically by methods previously described[12]. Magnetization measurements were carried out in pulsed fields with a sample extraction susceptibility coil, and in dc fields with an AFM piezoresistive cantilever in a bridge configuration[13,14]. The cantilever is placed at the center of the magnetic field so that the response is predominantly from torque, and not gradient forces. (A Meissner-type measurement was carried out independently at 1.7 K with a spherical lead sample to obtain an approximate calibration of the cantilever sensitivity.) Magnetothermopower studies were done in dc fields using a low frequency digital method[15]. The pressure measurements were electrical transport in an interplanar, 4 terminal configuration using gold wires and graphite paint. For pulsed field studies, a low pressure grease encapsulation method was employed, and in dc fields a standard BeCu pressure clamp was used. Both pressure methods were independently calibrated with conventional pressure sensors. Measurements were carried out at the National High Magnetic Field Laboratory: pulsed field (50 T - 20 ms) experiments were done at Los Alamos, and dc fields (45 T Hybrid) were used in Tallahassee.

**Results**

We first consider the MI transition as seen in electrical transport measurements under pressure and in magnetic field direction dependent MR studies. Comparisons are made between the pyrazino and the EDO compounds. This is followed by a description of magnetization measurements taken with both cantilever torque (dc field) and extraction (pulsed field) methods. Finally, we present the temperature dependence of the high field thermopower which reveals a gap opening for fields above the MI transition.

**A. Pressure dependence of the MI transition in dc magnetic fields**

To explore the effects of lattice constant changes on the MI transition, pressure dependent MR studies using a BeCu clamped cell were carried out simultaneously on three samples, $\tau(r)$-AuBr$_2$, $\tau$-AuBr$_2$, and $\tau$-EDO-AuBr$_2$, as shown in Figure 2. Measurements for the "finger tight" configuration of the pressure cell showed behavior very similar to previous studies. Here the samples are encapsulated in the pressure fluid, but there is no further compression of the fluid at room temperature. Hence the pressure at low temperatures is nearly zero, but indeterminate, i.e. $P = 0 + \varepsilon$. The MI transition typical of all pyrazino compounds was observed in $\tau(r)$-AuBr$_2$ and $\tau$-AuBr$_2$, as was the large background MR in the $\tau$-EDO-AuBr$_2$ sample above 30 T. SdH oscillations were also seen in all three materials. However, under a pressure of only 1 kbar, the MI transition in the pyrazino was completely suppressed, and SdH oscillations persisted to the highest fields. Likewise, the MR background in the EDO sample was dramatically reduced. (The MR background will be discussed in more detail below.) Note the low field hysteresis of the negative MR persists (inset, Fig. 2b), even at 5kbar, well above the threshold pressure needed to suppress the MI transition. Under these same conditions changes in the SdH frequencies were observed where τ-AuBr$_2$ showed a 7% increase in FFT frequency to 540 T. For convenience, we provide a summary of SdH frequencies and associated effective masses for these materials in Table 1.

The racemic material exhibited SdH oscillations for small pressure (0+ε kbar) which previously had not been observed in this variant of the pyrazino family in dc or pulsed field MR under ambient pressure (figure 2a). The Dingle temperature of the τ-AuBr$_2$ sample is lower than for the racemic τ(r)-AuBr$_2$, which is consistent with the more disordered nature of the racemic material. Using the Dingle temperature to estimate the relaxation time, $\tau$, ($T_D = \frac{h}{4\pi^2 k_B \tau}$) and the fundamental oscillation frequency (F$_3$) to estimate the Fermi velocity, $v_F$, we obtain a mean free path of 170 Å at T = 0.5K for the racemic sample. From the simple Drude picture, the zero field in-plane resistance yields a mean free path of the order of the unit cell (10 Å). Hence, although the temperature dependent resistivity suggests a "bad metal", the quantum oscillations show metallic character.

**B. Pulsed magnetic field studies – pressure effects in the pyrazino compound**

The high sensitivity of the MI transition to pressure, which is completely suppressed by 1 kbar, is difficult to explore with the standard pressure clamp system. Therefore a grease encapsulation method was used to apply a small, measurable pressure below 1 kbar. (This also avoided eddy current problems associated with BeCu clamped pressure cells that arise in pulsed fields.) We estimated the pressure at helium temperatures to be approximately 50 bar, based on independent calibrations with a semiconductor pressure gauge [16]). The MR was studied in a 50 T pulsed magnet, as shown in Fig. 3. The sample, first measured without the grease, showed the characteristic abrupt change at the MI transition accompanied by transients in the signal, as previously reported. [1] When the sample was measured under identical conditions, but with the grease pressure, the threshold field increased at the approximate rate of ≈ 43 T/kbar. With pressure, the transient effects were reduced, indicating a corresponding decrease in the insulating behavior. The low pressure study therefore confirmed the extreme sensitivity of the MI transition to pressure. By linear extrapolation at 1 kbar, $B_{MI}$ would be beyond 80 T.

**C. Pulsed magnetic field studies – angular dependent effects in the EDO compound**

Although there is no dramatic MI behavior in the EDO material, there is a background MR that appears above 30 T. Since, in the pryazino compounds the MI transition is nearly independent of field orientation [1], we carried out angular dependent MR measurements on the EDO system to examine the behavior of this background. Our results are shown in Fig. 4a for pulsed magnetic fields at low temperatures for different field orientations. We find that the background MR above 30 T is present in the B//c ($\theta = 0°$, where $\theta$ is the angle between the applied field and c-axis) data, and it remains, even for the B//ab-plane ($\theta = 90°$) orientation where no SdH oscillations appear. This indicates that there is a contribution to the MR in the EDO material that is independent of FS topology. The background MR increases significantly (although not as dramatically as the MI transition in the pyrazino systems) above 30 T. Since the SdH oscillation amplitude is proportional to the background MR, in Fig. 4b we have plotted the same results vs. inverse perpendicular field (B cos($\theta$)), where we have normalized the $\theta = 0°$, $30°$, and $60°$ data by dividing out the $\theta = 90°$ background MR. For comparison, we also show the $\theta = 0°$ data in figure 4b without normalization. If we focus on the low frequency (50 T) SdH oscillation, it is clear that its amplitude above 30 T deviates significantly from standard Lifshitz-Kosevich theory, since the amplitude is much larger than the exp(-T/B) dependence. However, upon normalization with the background MR, the final oscillation falls into the exp(-T/B) description. Hence we conclude from this study that there is an isotropic background MR which rises significantly above 30 T in the EDO system which is not part of the SdH signal.

**D. Magnetization**

To test the thermodynamic nature of the MI transition in the pyrazino materials, we carried out two complementary studies of the magnetization. In Figure 5 we present the torque magnetization signal for a τ-(P-(r)-DMEDT-TTF)$_2$(AuBr$_2$)$_{1+y}$ single crystal at 0.5 K mounted on an AFM-type cantilever. The overall change in the background signal with field is characteristic of the sample+cantilever response, which can change sign with different orientations in field[17]. However, the behavior of the hysteretic envelope above 35 T remains the same for all measurements. This behavior, shown in the Fig. 5 inset, is the signature of the MI transition in the magnetization. Of note, within the magnetization envelope that defines the hysteretic MI transition region, the signal appears reversible for increasing and decreasing fields. Since cantilever measurements are well suited for the measurement of the relative, but not the absolute moment, we have further quantified this change in magnetization by employing the extraction method in pulsed magnetic fields at 0.46 K, as shown in Fig. 6. Here the difference in the integrated sample-in and sample-out signals is shown vs. the magnetic field up-sweep. The magnetization, which increases linearly with magnetic field (paramagnetic), shows a slight increase in slope above 30 T. This further rise in magnetization is indicated as the deviation from linearity plotted against the right scale. The sizes of the total magnetization signals are in reasonable agreement, even with the difficulties in obtaining an absolute calibration for the very small samples on the cantilever and the correspondingly small moment changes observed.

The signature of the field-induced MI transition was further explored through the angular and temperature dependence of the magnetization in dc fields. In Fig. 7a we show the total magnetization for the field perpendicular to the conducting layers (B//c-axis) and in Fig. 7b parallel to the conducting layers (B//ab-planes). To mark the hysteretic transition field thresholds (B$_{MI}$-up and B$_{MI}$-down) we have used the peaks in the derivative of the signal, as indicated in Fig. 7a and b, and a complete plot of the angular dependence of the threshold fields is shown in Fig. 7c. The temperature dependence of the magnetization at the MI transition is shown in Fig. 8a where we have separated the up and down sweeps to highlight the different behavior of the two branches. We have used the derivatives of the signals (as in Fig. 7) to obtain the threshold fields, which are presented in Fig. 8b vs. in a T–B phase diagram. The low temperature appearance of the MI phase diagram is in good agreement with that obtained by transport, skin depth, and magnetocaloric measurements. This includes the behavior of the high field phase boundary which has a negative dT/dB slope for T → 0.

**E. Magnetothermopower study of the MI transition**

Magnetothermopower measurements for τ-(P-(*S,S*)-DMEDT-TTF)$_2$(AuI$_2$)$_{1+y}$ were carried out in constant dc fields vs. temperature as shown in Fig. 9. The inset of Fig. 9 shows the zero field temperature dependence. τ-AuI$_2$ is isostructural to the other pyrazino compounds and exhibits the same high field insulating state.[1]

We first discuss the zero field thermopower data, where several points should be noted. i) The sign of the thermopower is negative, which is consistent with the Hall effect measurement results on τ-AuBr$_2$ [10,18]. ii) The temperature dependence shows a broad maximum at around 50 K below which thermopower shows normal metallic behavior, i.e. decreases to zero almost linearly. This behavior is often observed in the in-plane thermopower of many two-dimensional organic conductors [19]. Merino *et al*. suggested that the peak in S(T) can be explained by the destruction of the Fermi liquid quasiparticles, resulting in a "bad" metal [20]. In this model, a smooth crossover from coherent Fermi liquid excitations at low temperatures to incoherent excitations at high temperatures leads to a non-monotonic temperature dependence in transport properties. The slope of metallic thermopower region of τ-AuI$_2$ is large compared to other organic conductors [20], indicating a small bandwidth, or large density of states at the Fermi level. We will return to this important point in the discussion section. The ratio of Coulomb repulsion to small measured bandwidth (U/W) would be large, revealing a tendency towards an insulating state, as seen in MR measurements. iii) Recalling the anomalous upturn in the resistance of the τ-AuBr$_2$ and τ-AuI$_2$ compounds [3,11] for T→0, the thermopower in the same region shows no anomalous behavior. Since thermopower is extremely sensitive to changes of electronic properties, but is rather insensitive to disorder, it is likely that anomalous behavior observed in the zero-field resistance measurement may be due to disorder inherent in this material.

In the magnetothermopower, the MI transition for B > B$_{MI}$ is clearly seen as an abrupt increase in S below the transition temperature T$_{MI}$. Upon further cooling, the thermopower decreases again, showing a broad peak. At lower temperatures, it was impossible to measure the thermopower signal since the sample impedance was too high and exceeded the range of the nanovoltmeters.

If we focus the thermopower behavior just below $T_{MI}$, the increase of thermopower can be explained as the opening of an energy gap in the insulating state. In Fig. 10, the data are presented to show the slope of S vs 1/T below $T_{MI}$, which is a measure of the activation energy gap, $E_g$. Since S(T) is proportional to the temperature dependence of the transport conductivity σ through ln(σ(T)), S(T) can be expressed as $S(T) \sim E_g / T$ and $E_g$, determined in this manner, in shown in the inset of Fig. 10. The energy gap opens at around B = 42 T and it increases with magnetic field.

The broad peak in thermopower below $T_{MI}$ is also been observed in other organic conductors [21], where the MI transition temperature is high enough so that the lower temperature behavior can be observed. This behavior may be due to carriers from impurities or defect levels that become important when the band carrier density decreases below the impurity/defect density.

**Discussion**

The field-induced MI transition is a general feature of the pyrazino τ-phase materials. All measurements to date, including electrical transport, skin depth, magnetocaloric effect, magnetization, susceptibility, and thermopower show that for many different pyrazino samples, there is a bulk, field-induced, hysteretic transition from a metallic state to an insulating state. Although the temperature dependence of the phase boundary is similar to that seen in the field-induced spin density wave (FISDW) states in the Bechgaard salts, there is little evidence that the MI transition involves FS nesting since the MI transition is only slightly dependent on field direction. Moreover, in the FISDW phases the conductivity remains finite, whereas in the pyrazino τ-phase materials it vanishes along all crystallographic directions [1]. (Ie., there is no evidence for imperfectly nested FS pockets remaining, as is generally the case in a FISDW system.) In this discussion, we consider several possible mechanisms that could either lead to, or strongly influence the field-induced MI transition.

The present work reveals that the MI transition is thermodynamic, and from the value of the magnetization change, we estimate that the transition involves an itinerant-to-localized transformation of about $10^{17}$ electrons/cm$^3$. Indeed, this change in carrier concentration is not far from that involved in the MI transition in, for instance, doped-semiconductor systems. The thermopower measurement further confirms the thermodynamic nature of the transition, since it shows that a field dependent energy gap opens above the MI transition.

In these materials, field-induced changes in electronic structure of order 10 meV or less could cause a catastrophic change, including a gap opening at the Fermi level. The tight binding electronic structure indicates that the bandwidth below the Fermi energy (see Fig. 1) is relatively shallow, of order 10 meV, and the band is very flat. Zero field thermopower measurements are consistent with this narrow bandwidth. The thermopower measurement also sets the relevant energy scale for the MI transition: the MI activation energy is field dependent, and reaches $E_a$ = 3 meV by 45 T (note $\Delta = 2E_a$). The other energy scales include the Fermi energy, the Zeeman band splitting, the Landau spectrum, and broadening due to disorder. If we consider a field of 40 T, and the largest ($F_3$) orbit, then $g\mu_B B = 4.6$ meV, $(\frac{\hbar eB}{m^*}) = 0.6$ meV, $E_F = \frac{\hbar eF}{m^*} = 8.4$ meV, and $T_D = 0.1$ meV, respectively. Due to the smaller values, the Landau gap and disorder probably do not strongly affect the MI transition. Rather, the band width, $E_F$, $E_a$, and the Zeeman energy appear to be the competing energy scales.

A simple model which would produce a field dependent energy gap is where the flat conduction band shifts relative to the Fermi level with increasing field. However, although the spin-down band will increase, the spin-up band will decrease in energy, and this cannot produce an insulating state if the Fermi level remains fixed. Experimentally we do not see evidence for spin split Landau levels, nor is there any indication of a change in the SdH frequency at the entry into the MI phase. (Even a careful investigation of the SdH structure on the rapidly rising resistance signals at the MI transition, in Fig. 2a for instance, shows that it is not shifted in frequency or phase from the SdH waveform in the lower field fully metallic phase.)

There are three other important factors in the behavior of the MI transition. The first is that it is highly suppressed in the EDO materials. The second is that the MI transition is hysteretic, indicating coupling between the electronic and lattice structures.

The third is the very high sensitivity of the MI transition to pressure. The pyrazino (N-N) locations in the asymmetric donor are not chalcogenides, and their molecular overlap with the adjacent S positions is not as strong as the corresponding EDO (O-O) system (i.e., the oxygen sites provide an additional electron pair orbital that the nitrogen sites do not). The asymmetric donor molecules are arranged in a square lattice, where each nearest-neighbor donors have opposite orientations. Therefore, the orbitals in the nitrogen (or oxygen) ring of one donor overlap with the orbitals in the sulfur rings of the neighboring donor, and these overlaps result in the main transfer integrals that give rise to the band structure. Hence, it is possible that the N-S coupling is susceptible to perturbation in high magnetic fields. Since this model depends on atomic or molecular orbitals, and not the FS topology, the field interaction could be relatively isotropic. If the magnetic field altered the coordination near or at the N-S coupling sites slightly, it would change the electronic structure, and it would be coupled to the lattice. Two nearly equivalent molecular configurations could give rise to the hysteretic behavior. The apparent reversibility within the hysteresis loops in Fig. 6, coupled with the widening of the hysteresis region at low temperatures, suggests a double-well process which is thermally activated. Previous magnetocaloric studies are also indicative of this kind of two-state switching, and also show the widening of the hysteretic region at low temperatures. The high sensitivity of the MI transition to pressure further supports the role the lattice plays in the close proximity of the MI and metallic phases.

The observation of a field-induced MI transition in complex organic materials is not without precedent. The Q2D organic superconductor, κ–(BEDT-TTF)$_2$Cu[C(NC)$_2$]Cl, also shows a field-induced MI transition in MR. The unit cell involves two donor layers, and the anion structure is sometimes referred to as "polymeric" [22]. Moreover, in this class of κ-(ET) salts, there is variable configurational order in the ethylene groups that affects both the superconducting transition and the appearance of SdH oscillations [23-25]. The non-metallic state measured in κ–(BEDT-TTF)$_2$Cu[C(NC)$_2$]Cl has been described in terms of loosely bound molecules that are susceptible to small perturbations of external parameters [26]. Given that the tau-phase materials have a four donor layer repeat unit cell, combined with a complex, non-stoichiometric anion structure, a similar sensitivity to perturbations can be expected.

The coupling of the magnetic field to the electronic structure may involve an enhanced susceptibility. Previously, Arita *et al.* [Ref. 8] have considered the flat band character of the electronic structure where it is argued that the magnetic susceptibility is enhanced due to the flatness of the bands along $k_x$ and $k_y$. Tight binding band structure calculations where the Zeeman and diamagnetic effects of high magnetic fields on the molecular orbitals would be instructive, since the field dependence of the electronic structure could be estimated. Likewise, structural probes including Raman, ultrasonics, or even X-ray diffraction would be very useful to gauge the role of the lattice at the transition.

**Acknowledgements**

This work is supported by NSF-DMR 02-03532, and the NHMFL is supported by a contractual agreement between the NSF and the State of Florida. DG is supported through a NSF GK-12 Fellowship.

**Figures and Captions**

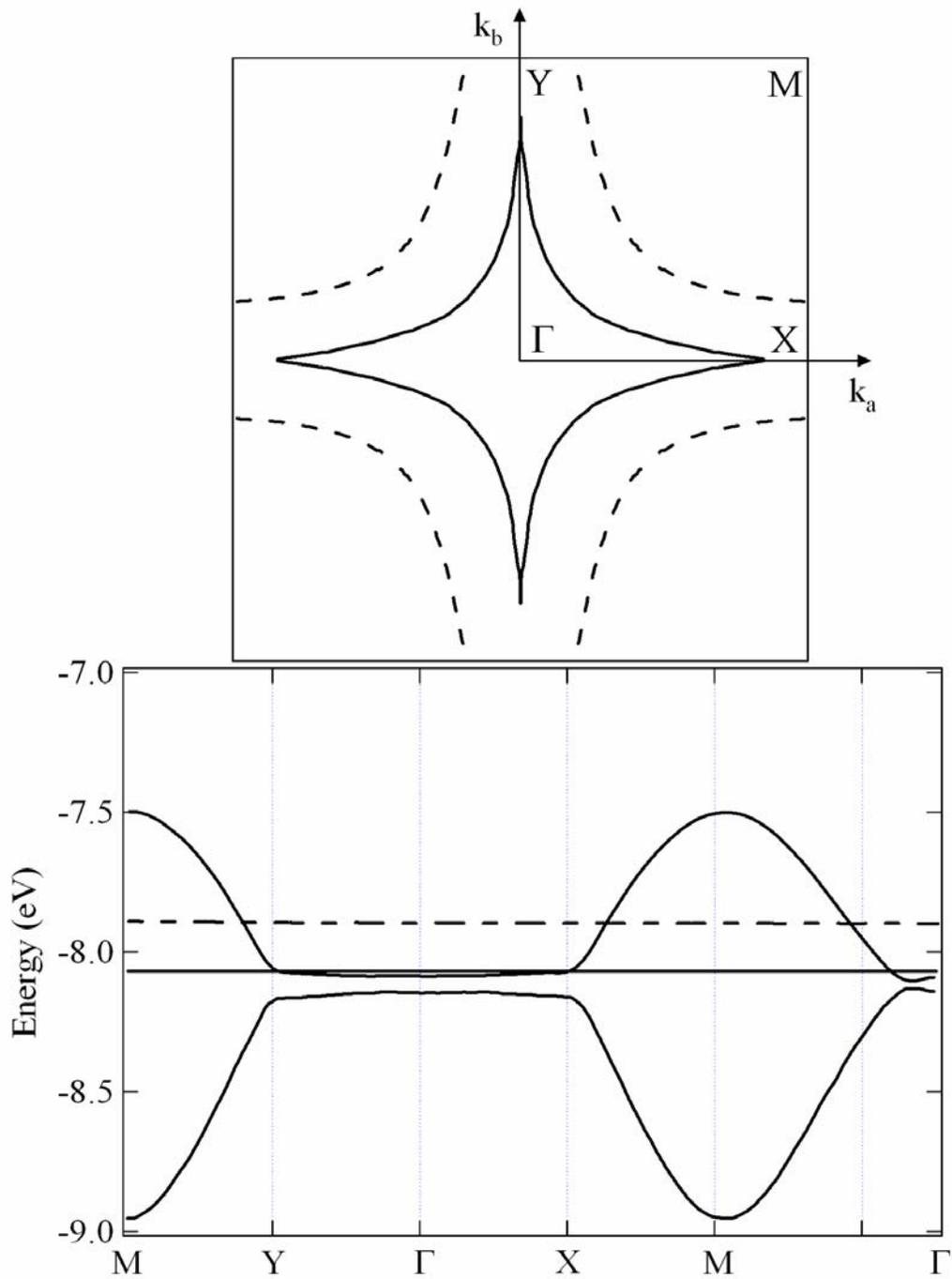

Figure 1.  a) Star-shaped FS and b) calculated band structure of τ-(P-(S,S)-DMEDT-TTF)$_2$(AuBr$_2$)$_{1+y}$. (Drawn after Ref. 4) The Fermi level lies very near to the band edge. In Figs. 1a and 1b: the dashed line represents y = 0 and solid line represents y ~ 0.75.

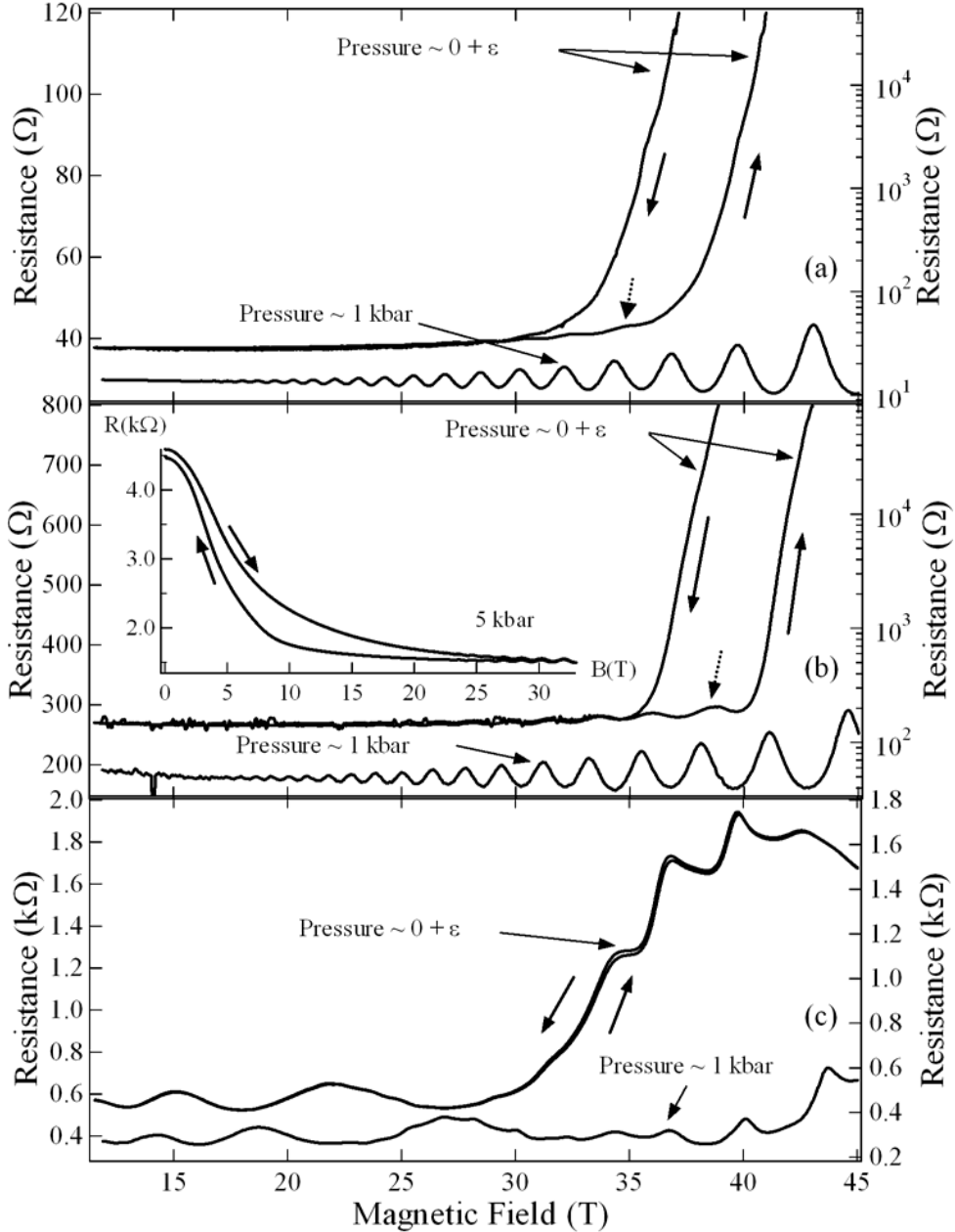

Figure 2.  Interplane resistance ($R_{zz}$) at T = 0.5K for (a) $\tau(r)$-AuBr$_2$, (b) $\tau$-AuBr$_2$ and (c) $\tau$-EDO-AuBr$_2$. Small pressure ε, left axis resistance; 1 kbar, right axis resistance. The field-induced MI transition in completely suppressed by 1 kbar, and the SdH oscillations indicate a well defined metallic state to the maximum field. The large background MR in the case of $\tau$-EDO-AuBr$_2$ is also suppressed by 1 kbar of pressure. Inset 2b: At lower fields the large hysteresis, characteristic of the $\tau$-phase materials is observed, even under 5 kbar of pressure.

| Pressure | Frequency | $\tau$-$(r)$AuBr$_2$ | $\tau$-AuBr$_2$ | $\tau$-EDO-AuBr$_2$ |
|---|---|---|---|---|
| $\varepsilon$ | $F_1$ | ----- | ----- | 47.3 |
| 1 kbar | $F_1$ | ----- | ----- | 58 (1.97) |
| $\varepsilon$ | $F_2$ | ----- | 166 | 175 |
| 1 kbar | $F_2$ | 185 (4.75) | 185 (4.45) | 200 (2.14) |
| $\varepsilon$ | $F_3$ | 497 | 505 | 465 |
| 1 kbar | $F_3$ | 494 (7.66) | 507 (7.80) | 484 (3.94) |
| 5 kbar | $F_3$ | ----- | 537 | ----- |

Table 1.   Summary of the Shubnikov-de Haas oscillation frequencies for $\tau(r)$-AuBr$_2$, $\tau$-AuBr$_2$ and $\tau$-EDO-AuBr$_2$. Where the temperature dependence of the SdH was measured, the effective mass parameter, m*/m$_0$, is given in parentheses after the SdH frequency.

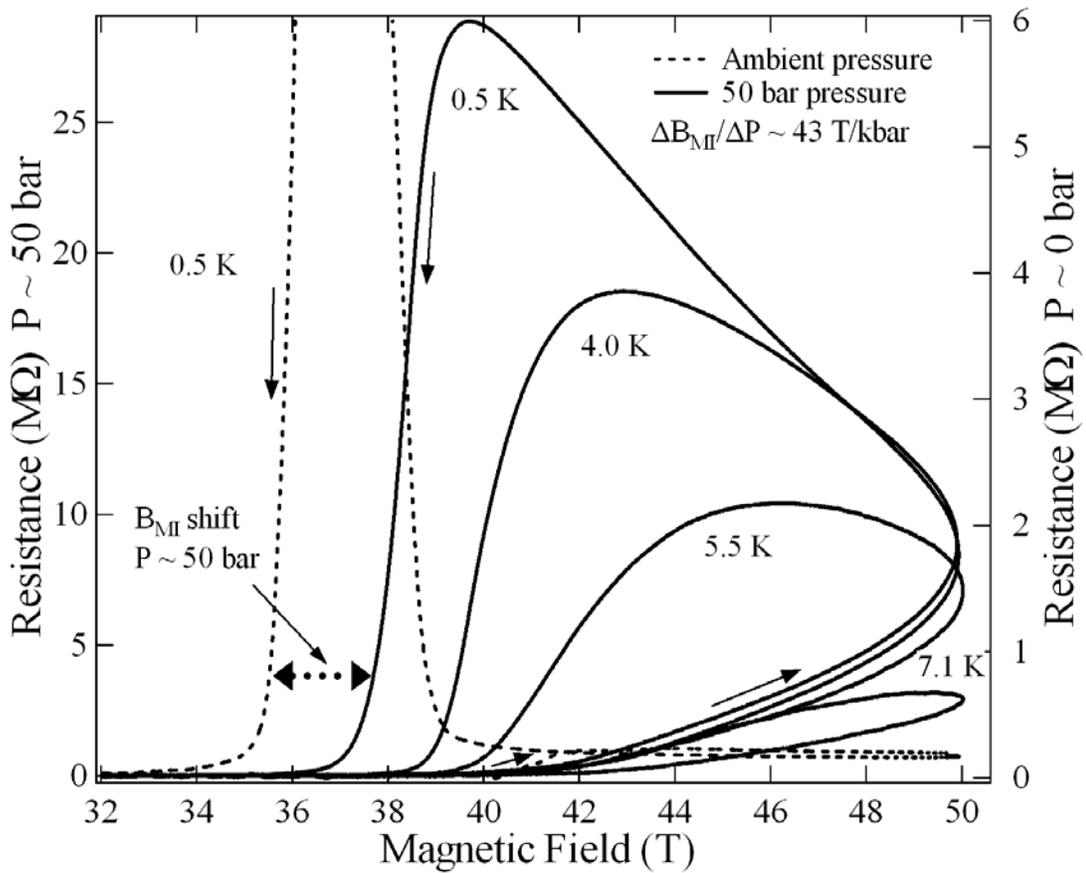

Figure 3. Low pressure study of the field-induced MI transition in the τ(r)-AuBr$_2$ material in pulsed fields at 0.5 K. Dashed line: ambient pressure signal, solid lines: 50 bar pressure signal for different temperatures. The thin, solid arrows represent the direction of the field sweep while the thick, dotted arrow shows the pressure dependent shift for the down-sweep.

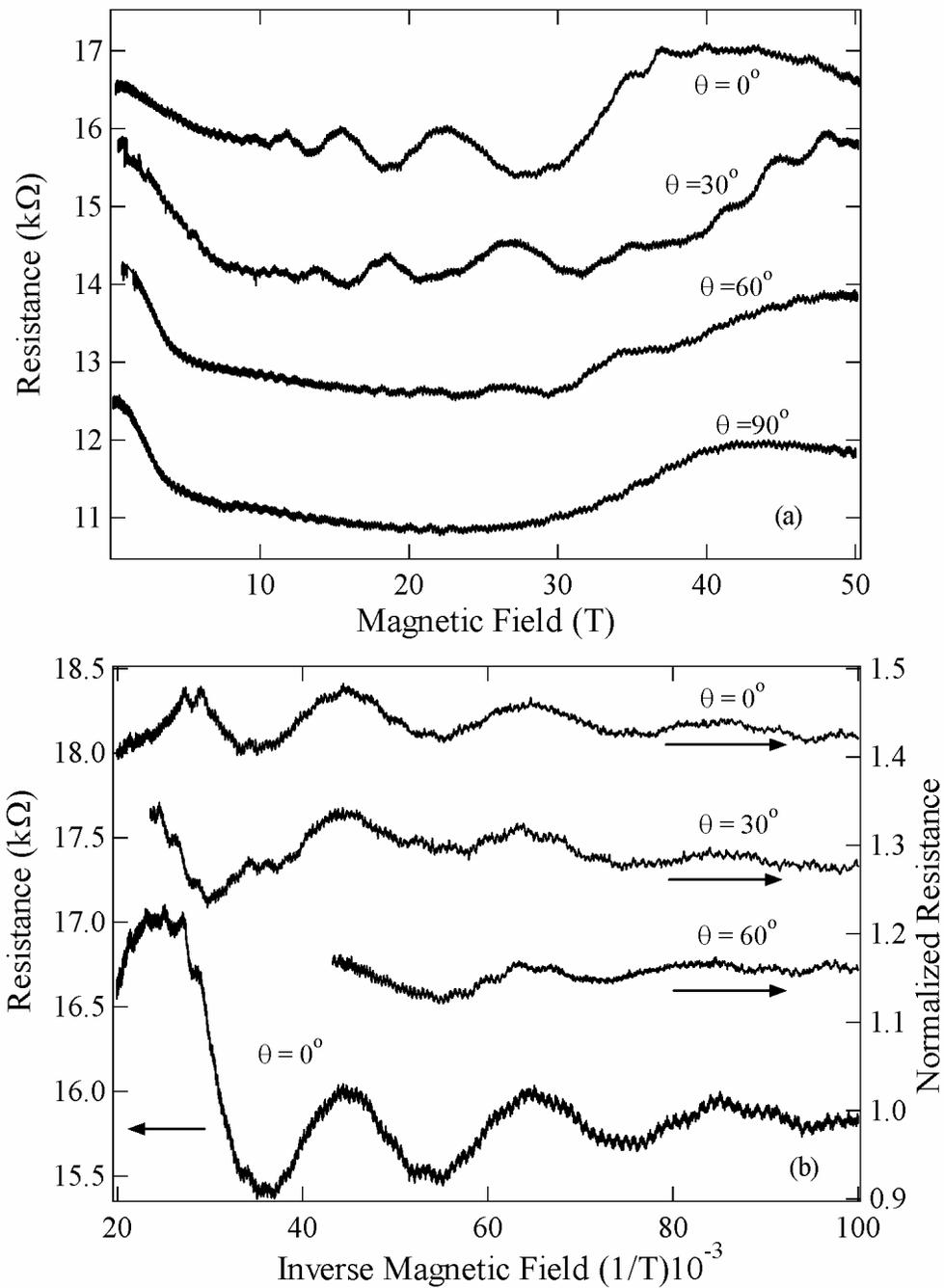

Figure 4. The angular dependent MR of the τ-EDO-AuBr$_2$ system at 0.5 K in pulsed fields. All traces are offset for clarity. a) MR data for different field orientations. b) Left axis: 0° data without normalization vs. inverse field. Right axis: Oscillatory MR vs. 1/Bcos(θ) normalized to the background MR from the 90° data in panel (a). See text for discussion.

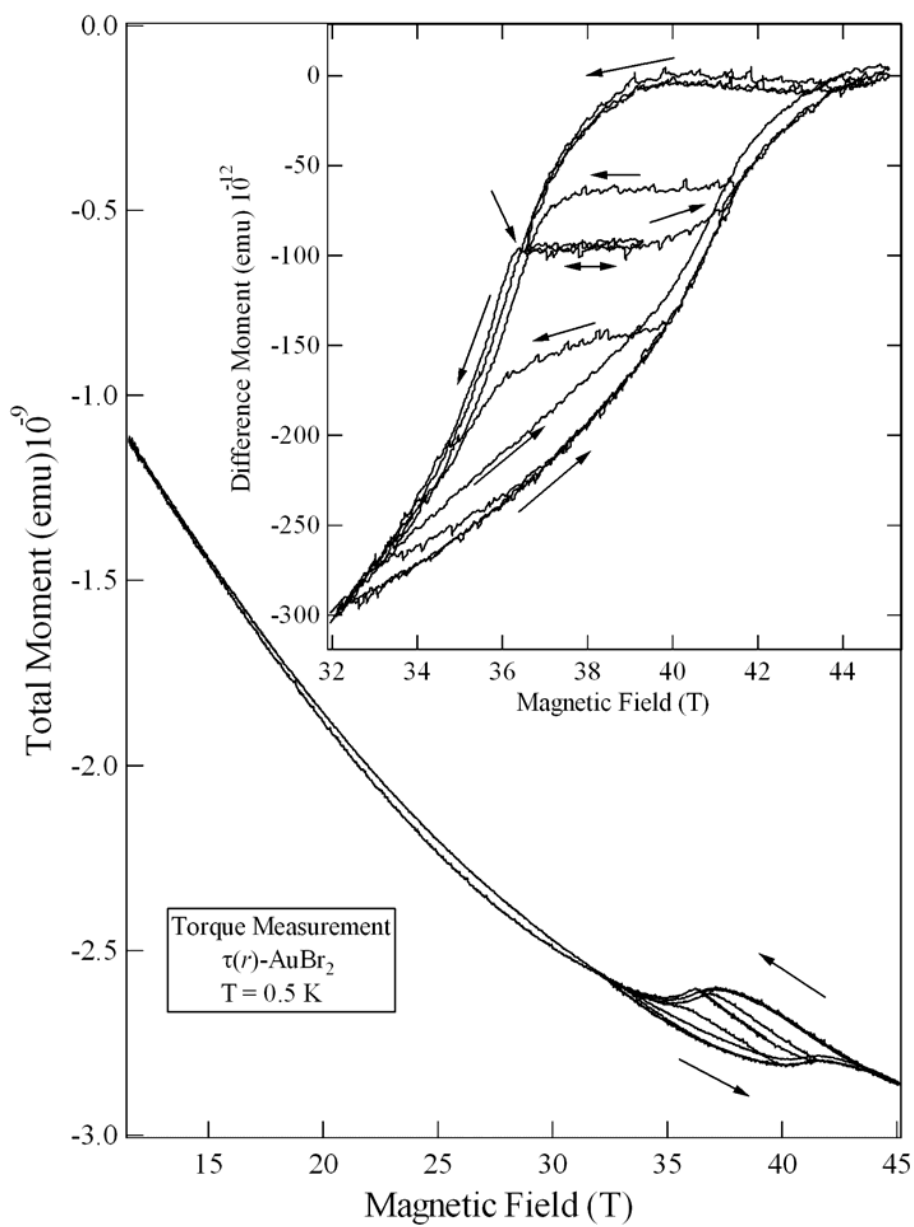

Figure 5. Cantilever torque magnetization signal from τ(*r*)-AuBr$_2$ at 0.5 K. The overall background signal depends on the orientation of the cantilever with respect to field. The inset shows detailed behavior of the hysteresis in the magnetization in the field-induced MI transition region. The arrows indicate the direction of the field sweep. The double arrow indicates reversible behavior between the upper and lower threshold limits.

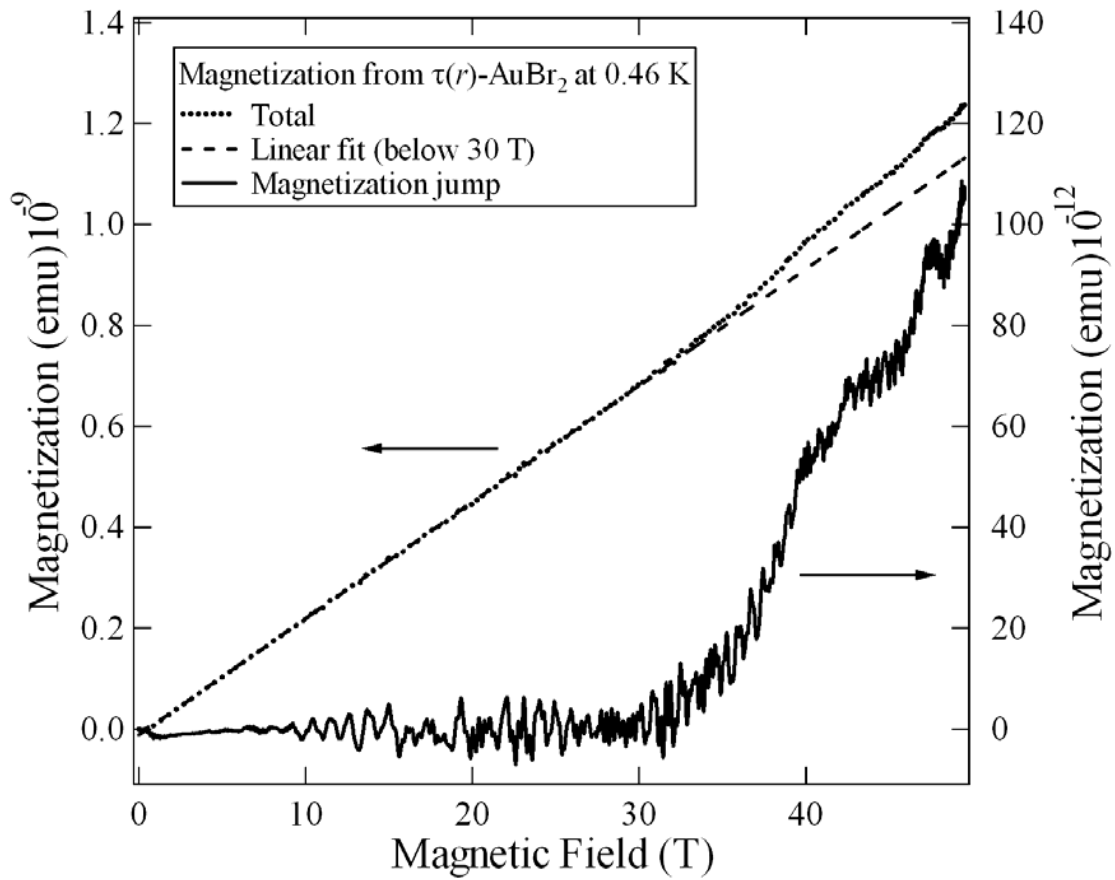

Figure 6. Magnetization of τ(*r*)-AuBr$_2$ at 0.46 K from a sample extraction susceptibility coil system in a pulsed field magnet.

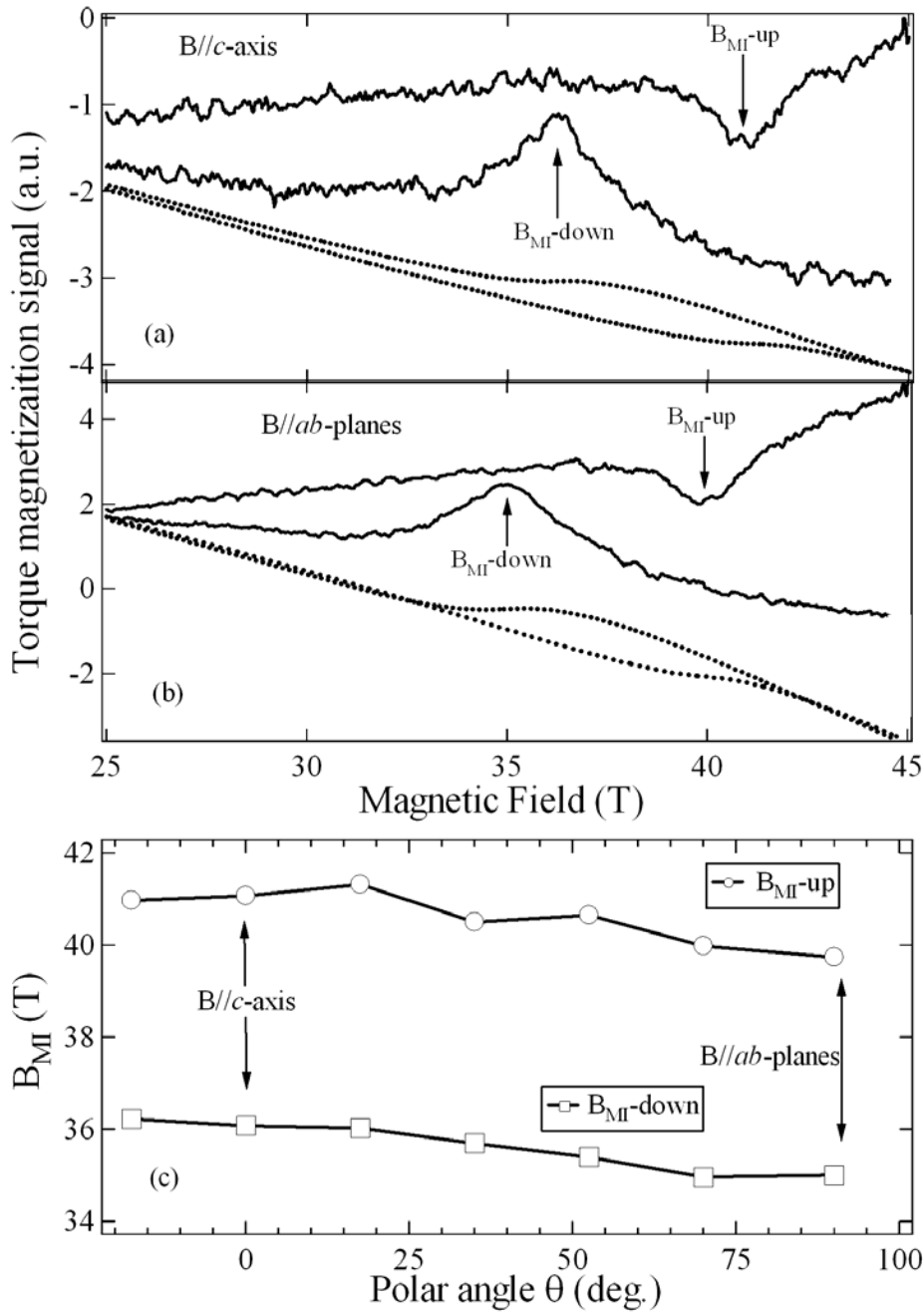

Figure 7. Cantilever torque magnetization signal from τ(r)-AuBr$_2$ at 0.5 K vs. field orientation. Threshold fields are determined from the peak in the derivative of the signal at the hysteretic field-induced MI transitions. a) B//*c*-axis: magnetization signal (dashed lines), derivative of signal (solid lines). b) B//*ab*-planes: magnetization signal (dashed lines), derivative of signal (solid lines). c) Field orientation dependence of hysteretic threshold fields B$_{MI}$-up and B$_{MI}$-down for full angular range.

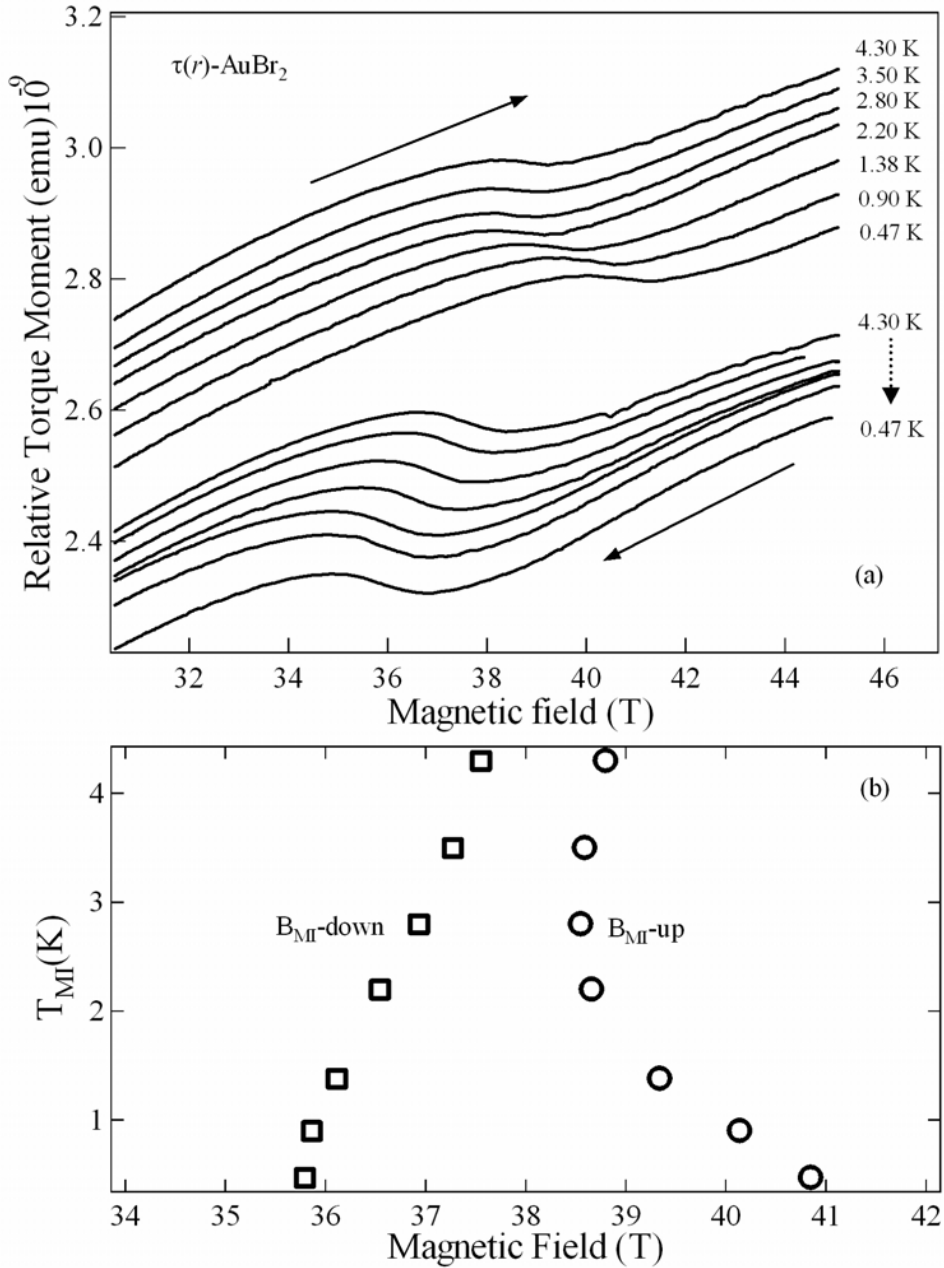

Figure 8. Cantilever torque magnetization signal from τ(r)-AuBr$_2$ for B//c vs. temperature. a) The up-sweep and down-sweep traces for each temperature have been offset to show the temperature dependence of the threshold field regions for B$_{MI}$-up and B$_{MI}$-down respectively. Solid arrows indicate direction of field sweep. b) Summary of the phase diagram for T$_{MI}$ vs. B$_{MI}$. The threshold fields were determined from the peaks in the derivatives of the traces, as indicated in Fig. 7. The increase in B$_{MI}$-up at lower temperatures is consistent with previous transport and magnetocaloric studies.

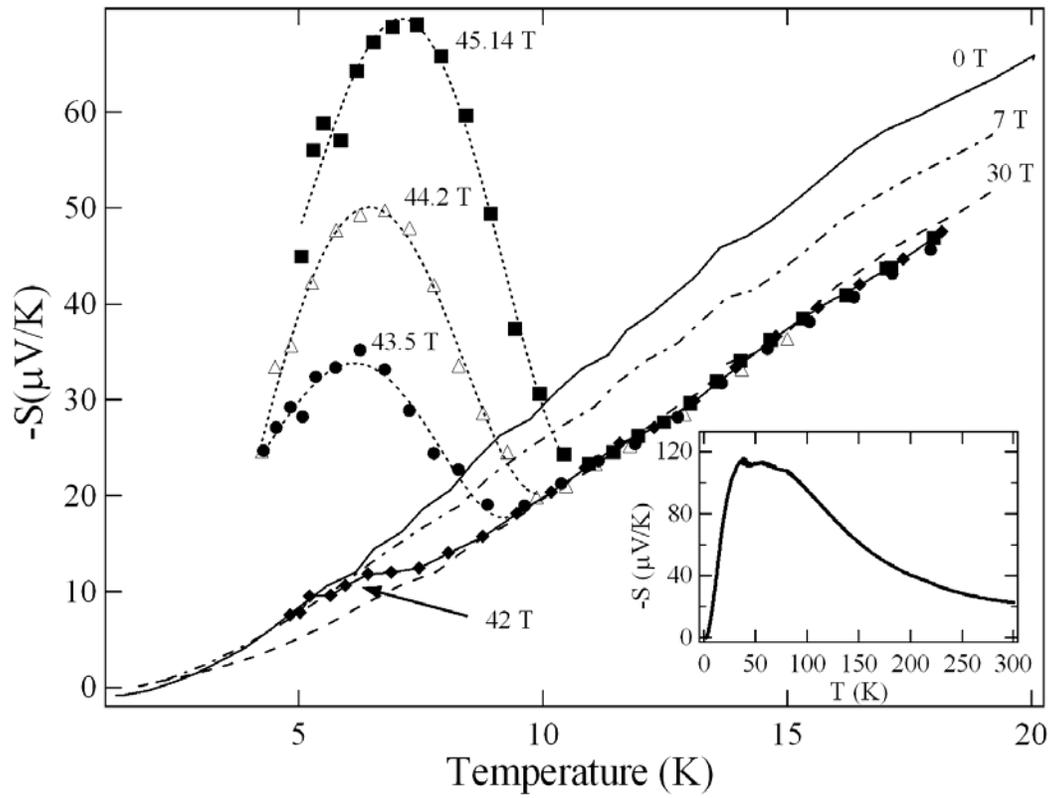

Figure 9. Magnetothermopower measurements for τ-AuI$_2$ carried out in constant dc fields vs. temperature in the region of the field-induced MI transition. Inset: zero-field thermopower vs. temperature.

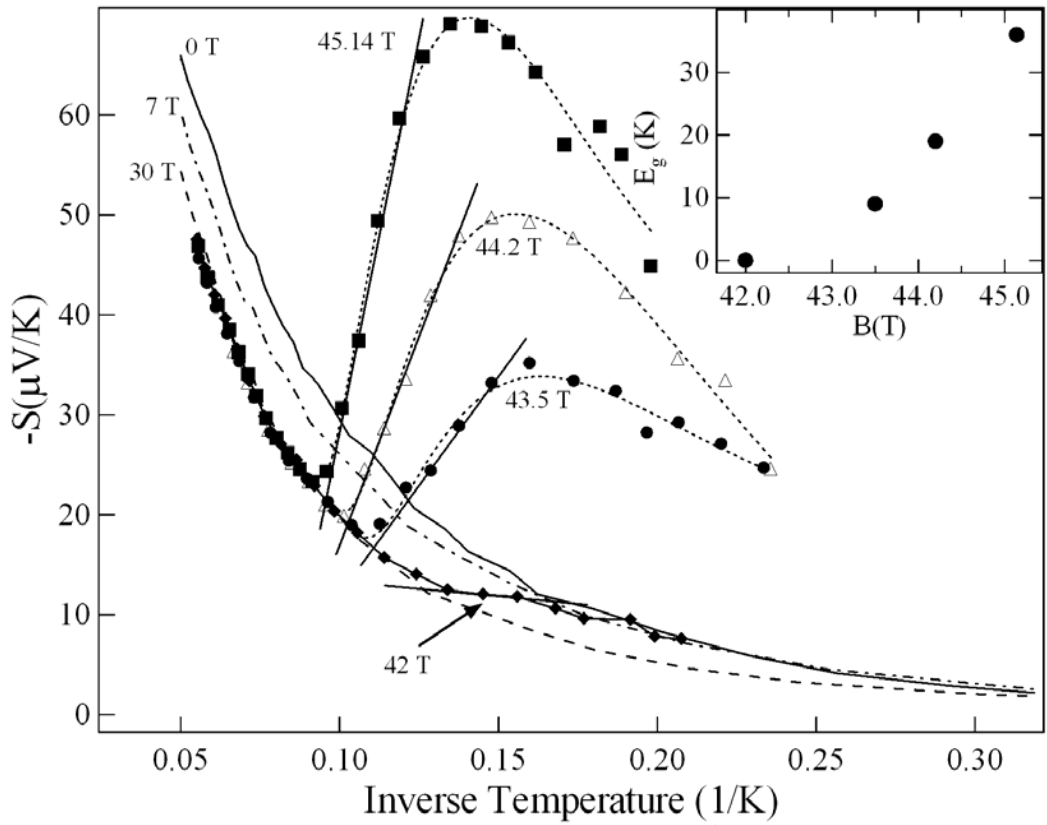

Figure 10. S(T) vs. 1/T representation of magnetothermopower measurements in the field-induced MI regime. Lines are fits to the activated behavior below the transition for each field. Inset: Field dependence of the activation energy.